\documentstyle[12pt]{article}
\begin{document}
\begin{center}
{\Large \bf
Indeterministic Quantum Gravity} \\[1.5cm]
{\bf Vladimir S.~MASHKEVICH}\footnote {E-mail:
gezin@gluk.apc.org (c/o S.V.~Mashkevich)}  \\[1.4cm]
{\it Institute of Physics, National academy
of sciences of Ukraine \\
252028 Kiev, Ukraine} \\[1.4cm]
\vskip 1cm

{\large \bf Abstract}
\end{center}

A theory which claims to describe all the universe is
advanced. It unifies general relativity, quantum field
theory, and indeterministic conception.

Basic entities are: classical metric tensor $g$, cosmic
reference frame (including cosmic time $t$), operator $T$ of
energy-momentum tensor, Hamiltonian $H_t$, and state vector
$\Psi$. Dynamical equations are: the Einstein equation
$G[g]=(\Psi,T\Psi)$ ($G$ is the Einstein tensor), the
Heisenberg equation $dT/dt=i[H_t,T]$, and the condition
$H_t\Psi_t=\varepsilon_t\Psi_t$ arising from the cosmic energy
determinacy principle advanced in the theory. The last
equation describes quantum jump dynamics.

Quantum jumps lead to the instantaneous transferring of action
and information, which, however, neither violates the
causality principle, nor contradicts quantum field theory and
general relativity.

The cosmic energy determinacy principle implies the eternal
universe, i.e., the cyclic one without beginning and ending,
the minimal energy in every cycle being finite.

\newpage

\hspace*{7 cm}
\begin{minipage} [b] {8.3cm}
        Les plus riches cit\'es, les plus grands paysages, \\
        Jamais ne contenaient l'attrait myst\'erieux  \\
        De ceux que le hasard fait avec les nuages. \\
        Et toujours le d\'esir nous rendait soucieux! \\
\end{minipage}
\begin{flushright}
Charles Baudelaire \vspace*{0.8 cm}
\end{flushright}

\begin{flushleft}
\hspace*{0.5 cm} {\Large \bf Introduction}
\end{flushleft}

Current physics is based on two fundamental theories and one
fundamental conception and is confronted with two fundamental
problems. The theories are quantum field theory (QFT) and
general relativity (GR), the conception is quantum
indeterminism, the problems are quantum gravity and
indeterministic quantum theory. But as long as the problem of
a unified theory is raised, the global problem may be
formulated as that of constructing indeterministic quantum
gravity, i.e., a theory which would unify QFT and GR and
describe quantum jump dynamics. Such a theory would, naturally,
claim to describe all the universe.

The aim of this paper is to construct indeterministic quantum
gravity.

The usual approach to the problem of constructing a unified
physical theory reduces to unifying QFT and GR. Such a unified
theory is usually said about as quantum theory of gravitation,
or quantum gravity. There is voluminous literature devoted to
problems of this approach (see, e.g., [1-10]). But there
exists yet another fundamental problem---that of quantum
jump dynamics. This is the starting point of the approach
advanced in this paper. The reasoning is as follows.

In indeterministic dynamics [11], a quantum system being
initially in a pure state $\omega'=(\Psi',\cdot\Psi')$ passes,
in general, into a mixed state
$\omega''=\sum_{j\in J}w_j\omega_j, \omega_j=(\Psi_j,\cdot\Psi_j),
(\Psi_j,\Psi_{j'})=\delta_{jj'}$. The interpretation of the
latter is the following: "In reality", the system is with a
probability $w_j$ in a pure state $\omega_j$. Quotation marks
are put because the corresponding notion is primary. The
transition $\omega'\rightarrow\omega''$ is usually considered
as the result of the breaking of coherence between amplitudes
$\Psi_j$'s [12--14,11]. Notwithstanding great efforts, this
approach has not hit the mark. In this paper the aim will be
achieved in quite another way. Now, however, we focus
attention on the transition $\omega''\rightarrow\omega_j$,
which is the reduction of the mixed state to the pure one.
Here, a problem arises with the hierarchy of irreduction, i.e.,
the absence of reduction, or retaining the mixed state [15,16].
(Do not confuse with the von Neumann hierarchy related to the
problem of coherence breaking, i.e., the transformation of a
pure state into a mixed one.) There is vagueness concerning the
stage of time evolution on which the reduction occurs. The
problem of irreduction would be solved if indeterministic
dynamics were non-linear with respect to states.

A possibility to construct such a dynamics arises by taking
into account gravitation in connection with the Einstein
equation. The Einstein tensor $G[g]$ is non-linear with
respect to the metric $g$. Therefore the averaging of the
quantum energy-momentum tensor $T$ with respect to the mixed
state $\omega''$ is not equivalent to the averaging with
respect to the individual pure constituents $\omega_j$ with
the subsequent averaging of the corresponding $g_j$'s with
the probability distribution $\{ w_j, j\in J\} :
g''\neq \sum_jw_jg_j$. Thus, if one is based on the equation
$G=\omega (T)$ the problem of irreduction hierarchy will be
solved. In the usual terminology, such a quantum theory of
gravitation is called semiclassical.

The next and crucial point is the following. Denote $a(t)$ the
cosmic scale factor in the Robertson-Walker spacetime. In GR
the state of the matter---energy or energy density in the
Friedmann model---does not depend on $da/dt$, which is an
adiabatic effect. Correspondingly, we assume that the state
vector of matter should be an eigenvector of the Hamiltonian
$H_t$: $H_t\Psi_t=\varepsilon_t\Psi_t$. This may be called the
principle of cosmic energy determinacy. At branching, i.e.,
crossing and splitting points of energy levels, this principle
gives rise to quantum jumps. Thus quantum jump dynamics arises.

The cosmic time $t$ is determined through the metric $g$, which
is the third essential point.

Now indeterministic quantum gravity may be formulated. Basic
entities are: the classical metric tensor $g$, the operator $T$
of the energy-momentum tensor, and the state vector $\Psi$.
Through $g$, the cosmic time $t$, the corresponding cosmic space,
and the Hamiltonian $H_t$ are determined. Dynamical equations
are: the Einstein equation $G[g]=(\Psi,T\Psi)$, the Heisenberg
equation $dT/dt=i[H_t,T]$, and equations for $\Psi_t$ arising
from  the cosmic energy determinacy principle.

QFT and GR are limiting cases of the theory.

Quantum jumps lead to the instantaneous transferring of action
and information, which, however, neither violates the causality
principle, nor contradicts QFT and GR.

The time dependence of the particle energy is known for  the
Robertson-Walker spacetime. This dependence implies that
quantum jumps may be connected with particle transformations.
This opens a new avenue of attack on the problem of quantum
measurement.

The cosmic energy determinacy principle implies the eternal
universe, i.e., the cyclic one without beginning and ending,
the minimal energy in every cycle being finite.

\section{The problem of unified physical theory}

It is conventionally thought that current physics is founded
on two fundamental theories: QFT and GR. But QFT per se does
not include the conception of quantum indeterminism, so that
the latter should be related to the foundations of physics.
Thus any unified theory must incorporate the three fundamental
entities: QFT, GR, and quantum indeterminism. However, main
efforts are directed to unifying either QFT and GR or QFT and
quantum indeterminism. Had the efforts been successful, the
former direction would have resulted in quantum gravity, the
latter in indeterministic quantum theory. There exist, however,
no consistent theories of this kind.

\subsection{The problem of quantum gravity}

Both in GR and in QFT, there are, first of all, the objects of
two kinds: 1) a spacetime $(M,g,\nabla )$, i.e., a Lorentz
manifold on which the Levi-Civita connection $\nabla$ is
defined, where $M$ is a differentiable manifold; 2) a family
$F$ of material fields.

In QFT, a special case of the Lorentz manifold figures, namely
the Minkowski manifold $(R^4,\eta)$, where $\eta$ is the
Minkowski metric. In GR, material fields are classical (matter
and electromagnetic field). In QFT, material fields are quantum
(spinor and gauge fields). In both theories, material fields
are considered as those on the spacetime manifold $M$. The
effect of the material fields on the manifold is taken into
account in GR but not in QFT. In both theories, the metric $g$
is a classical (tensor) field. This is natural in GR. In QFT,
it is possible to avoid the problem of quantum description of
spacetime inasmuch as the effect of the quantum material fields
on the spacetime is not taken into account. In gravitation
theory, the effect of the material fields on the spacetime
$(M,g)$ should be taken into account. In quantum gravity, these
fields must be quantum. Thus, there arises the problem of
taking into account the effect of quantum material fields on
the manifold $(M,g)$, i.e., of describing $(M,g)$ in quantum
gravity.

In the usual approach, the problem is posed in such a way: what
and how is to be quantized in $(M,g,\nabla )$? Such an approach,
at first sight, seems to be natural. Indeed, in GR the effect
of matter on the spacetime is defined by the dynamic equation
of GR, i.e., the Einstein equation $G=T$, where $G$ is the
Einstein tensor, $T$ is the energy-momentum tensor. In QFT the
field $T$ is quantum, therefore if the Einstein equation were
directly transferred into quantum gravity, the field $G$ should
become quantum. But $G=G[g]$ is defined by the metric $g$, so
that, one would think, the latter should be quantized. And,
what is more, one might try to quantize the manifold $M$ itself.
However, as it will be clear from what follows, the situation
is by no means such obvious. Be it as it may, this approach has
not been crowned with success.

\subsection{The problem of indeterministic quantum theory}

In QFT, dynamics is deterministic, a dynamic transition being
described by a symmetry in the Schr\"odinger picture and by a
Jordan *-automorphism in the Heisenberg picture. The
GNS-construction reduces both the symmetry and the automorphism
to a unitary operator in a Hilbert space. In such a transition,
pure states turn into pure ones.

Quantum indeterminism manifests itself in quantum jumps, so
that the problem of indeterministic quantum theory is that of
indeterministic dynamics, i.e., quantum jump dynamics. A
quantum jump is the transition
$$\omega'\rightarrow\omega''=\sum_{j\in J}w_j\omega_j
\rightarrow\omega_j, j\in J,         \eqno{(1.2.1)}$$
where
$$\omega'=(\Psi',\cdot\Psi'), \omega_j=(\Psi_j,\cdot\Psi_j),
(\Psi_j,\Psi_{j'})=\delta_{jj'}.     \eqno{(1.2.2)}$$
The transition $\omega'\rightarrow\omega''$ is usually
considered as the result of the breaking of coherence between
the summands in
$$\Psi'=\sum_{j\in J}c_j\Psi_j, c_j=(\Psi_j,\Psi').
\eqno{(1.2.3)}$$
But no clear result has been obtained on this way.

\subsection{Indeterministic theory of gravitation}

In our papers [15,17], a single theory which unifies QFT,
quantum indeterminism, and GR is constructed. But in this
theory, indeterministic dynamics is based on the approach of
coherence breaking, which seems inadequate.

\section{Heuristic considerations and conceptions of the
theory}

{}From the previous discussion, it seems that the attempts to
unify both QFT with GR (without taking into account
indeterminism) and QFT with indeterminism (without gravity)
are inadequate. Then the only way that remains is to take into
account all the entries---QFT, GR, and indeterminism ---
from the outset.

\subsection{Irreduction hierarchy problem and semiclassical
theory of gravitation}

We begin with unifying QFT and GR, having taken into account
indeterminism.

In quantum indeterminism, the problem of von Neumann hierarchy,
or coherence hierarchy, figures. Usually, it is considered in
connection with the measurement problem. The essence of the
coherence hierarchy problem consists in that it is not known on
what stage the coherence of summands of a state amplitude
(1.2.3) should be considered to be broken. In the mathematical
part of QFT, i.e., in deterministic dynamics this problem
cannot be solved. It is the interpretational part of QFT where
it is possible to introduce coherence breaking, i.e., to cut
off the hierarchy.

In quantum indeterminism, there exists yet another hierarchy,
namely, that of irreduction. The essence of the corresponding
problem is that it is not clear on what stage of a dynamic
process the reduction, i.e., the transition
$\omega''\rightarrow\omega_j$ in (1.2.1) should be considered
to be realized. Often, it is done on the level of the observer
as it was done by von Neumann in the case of coherence
hierarchy. But unless one refers to the observer, one may
consider that reduction is not realized at all, i.e., the
system remains in the mixed state $\omega''$ all the time.

But the notion "in reality", or the concept of reality, is so
fundamental that it is greatly desirable that this notion
appear yet in the mathematical part of the theory. To achieve
this, an indeterministic theory in which dynamics would be
non-linear with respect to states is required. A possibility
to construct such a dynamics arises by taking into account
gravitation.

In semiclassical theory of gravitation (see, e.g., [1]),
material fields are described in a quantum way, and spacetime
is described classically. The dynamic equation corresponding
to the Einstein equation is
$$G[g]=\omega(T),  \eqno{(2.1.1)}$$
where $\omega$ is the state of material fields in the
Heisenberg picture. Thus, the metric $g$ is a classical tensor
field on the differentiable manifold $M$. The quantum field
$T=T[F]$ is expressed in terms of the family $F$ of quantum
material fields on the manifold $(M,g,\nabla)$.

The equation (2.1.1) is non-linear with respect to $g$. In the
case of the state
$$\omega=\sum_jw_j\omega_j,  \eqno{(2.1.2)}$$
the averaging on the right-hand side of the eq.~(2.1.1) with
respect to $\omega_j$ gives a metric $g_j$, the averaging with
respect to $\omega$ gives a metric $g$, and
$$g\neq \sum_jw_jg_j. \eqno{(2.1.3)}$$
Thus, indeterministic dynamics obtained in such a way would be
non-linear.

Let us advance some arguments for supporting semiclassical and,
by the same token, indeterministic theory of gravitation, the
latter being developed on the basis of the former.

In classical mechanics, one usually does not make a distinction
between the time dependence of an abstract observable in the
Hamilton picture and the change of its mean value. However, in
principle the distinction does exist. Accordingly, the
classical Einstein equation $G=T$ may be treated in the two
ways. But in quantum theory these two ways will not be
equivalent: The former results in the operator equation, the
latter leads to the equation for mean values. In the latter
case, it is possible to retain the classical description of
spacetime, which results in the eq.~(2.1.1).

\subsection{Adiabatic effect in GR and cosmic energy
determinacy principle}

In GR, in classical cosmology specifically, the state of
matter at a time $t$ depends on the state of spacetime at the
time $t$ only but not on prehistory. Namely, the energy density
depends only on the cosmic scale factor $a(t)$ but not on
$a(t'), t'<t.$ This is an adiabatic effect (in mechanical, not
thermodynamical meaning). What corresponds to this in quantum
mechanics is the adiabatic approximation for the time-dependent
Schr\"odinger equation with time-dependent Hamiltonian $H_t$.
Thus, we assume that the state vector $\Psi$ of matter should
be an eigenvector of the Hamiltonian:
$$H_t\Psi_t=\varepsilon_t\Psi_t.  \eqno{(2.2.1)}$$
This fundamental requirement will be called the principle of
cosmic energy determinacy. It plays a crucial role in the
theory being developed.

\subsection{Robertson-Walker spacetime and cosmic time
and space in the non-homogeneous universe}

The above discussion provides the existence of a preferred
universal time $t$. Consider this provision.

In the general case, the manifold $M$ in its own account has
no additional structure; specifically the Lorentz manifold is
$(M,g)$.

In the Robertson-Walker model, $M$ does have an additional
structure: it represents the direct product of two manifolds:
$$M=S\times T, M\ni q=(s,t), s\in S, t\in T. \eqno{(2.3.1)}$$
A 3-dimensional manifold $S$ is a space, a 1-dimensional one
$T$ is time. The structure (2.3.1) is analogous to that of the
Aristotelian manifold. In both cases, it is possible to speak
of an absolute (or cosmic) time $T$ and an absolute space $S$.
The manifold $S$ in the Aristotelian case is an affine space,
in the Robertson-Walker model it is such only with the flat
space. The manifold $T$ in the Aristotelian case is an affine
space isomorphic to the real axis, in the Robertson-Walker
model $T$ is a real axis interval.

Thus, in the homogeneous universe, there exists the cosmic
time. The homogeneous model being a good approximation, we
assume  the existence of a cosmic time in the non-homogeneous
universe, too.

\section{Formulation of the theory}

Now it is possible to formulate indeterministic quantum gravity
as a consistent theory.

\subsection{The universe as a physical system}

The universe $U$ as a physical system is a pair $(st,m)$, where
$st$ is spacetime, $m$ is matter. The spacetime is
$(M,g,\nabla)$, the matter is a family $F$ of quantum fields
on $M$.

The basic dynamic quantities are: The classical metric tensor
$g$, the operator $T$ of the energy-momentum tensor, and the
state vector $\Psi$ of the matter.

Dynamic equations are ones for these quantities. One of the
equations is (2.1.1), or $G[g]=(\Psi,T\Psi)$. To formulate
the other two equations it is necessary to introduce cosmic
time and space.

\subsection{Cosmic time and space. Hamiltonian}

Cosmic time may be introduced by considering eq.~(2.3.1) as the
trivial fibre bundle and extending this construction for the
non-homogeneous universe. But a simpler way is to employ the
concept of reference frame [18].

A reference frame $Q$ on a spacetime $(M,g,\nabla)$ is a vector
field each of whose integral curves is an observer, i.e.,
$g(Q,Q)=1$ and $Q$ is future pointing. Let $\omega$ be the
1-form physically equivalent to $Q$: $\omega=g(Q,\cdot)$. $Q$
is called proper time synchronizable iff $\omega=dt$. The
function $t$ on $M$ is called a proper time function for $Q$.
This function plays the role of cosmic time. Any level
hypersurface $S_t$ of the function $t$ may serve as a cosmic
space.

The Hamiltonian is $H_t=\int_{S_t}\mu(ds)T_{00}(s,t)$.

The second dynamic equation is the Heisenberg equation for
$T$: $dT/dt=i[H_t,T]$.

Some commutation relations for $T$ should be established.

\subsection{The time dependence of state vector}

We assume the universe to be spatially finite (closed
universe). Then the spectrum of $H_t$ is discrete, the same is
the set of branching (crossing and splitting) points of levels
of $H_t$.

Let $\Psi_t$ be an eigenvector belonging to a level between two
branching points, and let $P_t$ be the corresponding projector.
We have from the condition (2.2.1)
$$P_t\Psi_t=\Psi_t,  \eqno{(3.3.1)}$$
and
$$\Psi_{t+dt}=P_{t+dt}\Psi_t.  \eqno{(3.3.2)}$$
We shall see that
$$\|\Psi_{t+dt}\|=1.  \eqno{(3.3.3)}$$
{}From (3.3.1),(3.3.2) it follows
$$\frac{d\Psi_t}{dt}=\frac{dP_t}{dt}\Psi_t. \eqno{(3.3.4)}$$

{}From
$$(P_t+\frac{dP_t}{dt}dt)(P_t+\frac{dP_t}{dt}dt)=
P_t+\frac{dP_t}{dt}dt  \eqno{(3.3.5)}$$
we obtain
$$P_t\frac{dP_t}{dt}+\frac{dP_t}{dt}P_t=
\frac{dP_t}{dt},  \eqno{(3.3.6)}$$
whence
$$(\Psi_t,\frac{dP_t}{dt}\Psi_t)=0.  \eqno{(3.3.7)}$$
Now
$$(\Psi_{t+dt},\Psi_{t+dt})=([P_t+\frac{dP_t}{dt}dt]\Psi_t,
[P_t+\frac{dP_t}{dt}dt]\Psi_t)=
(\Psi_t,[P_t+\frac{dP_t}{dt}dt]\Psi_t)=1.  \eqno{(3.3.8)}$$

Thus, the continuous change of $\Psi_t$ is determined by
eq.~(3.3.4).

Now consider a branching point at $t=t_b$. Let
$$\Psi_{t_b}=\sum_kP_{k\,t_b+0}\Psi_{t_b}  \eqno{(3.3.9)}$$
where $k$ is the branch number. Then a quantum jump occurs:
$$\Psi_{t_b}\rightarrow\Psi_{k\,t_b+0}=
\frac{P_{k\,t_b+0}\Psi_{t_b}}{\|P_{k\,t_b+0}\Psi_{t_b}\|}
\eqno{(3.3.10)}$$
with the probability
$$w_k=(P_{k\,t_b+0}\Psi_{t_b},P_{k\,t_b+0}\Psi_{t_b})=
(\Psi_{t_b},P_{k\,t_b+0}\Psi_{t_b}).  \eqno{(3.3.11)}$$

\subsection{Dynamic equations}

Let us collect the results. The dynamic equations are:
$$G[g]=(\Psi,T\Psi);  \eqno{(3.4.1)}$$
$$\frac{dT}{dt}=i[H_t,T],  \eqno{(3.4.2a)}$$
$$\mbox{commutation relations for }T;  \eqno{(3.4.2b)}$$
$$\frac{d\Psi}{dt}=\frac{dP_t}{dt}\Psi, t\neq t_b,
\eqno{(3.4.3a)}$$
$$\Psi_{t_b}\rightarrow\Psi_{t_b+0\,k}
=\frac{P_{k\,t_b+0}\Psi_{t_b}}{\|P_{k\,t_b+0}\Psi_{t_b}\|},
w_k=(\Psi_{t_b},P_{k\,t_b+0}\Psi_{t_b});  \eqno{(3.4.3b)}$$
$$H_t=\int_{S_t}\mu(ds)T_{00}(s,t);  \eqno{(3.4.4)}$$
$$g(Q,Q)=1, g(Q,\cdot)=dt.  \eqno{(3.4.5)}$$

Equation (3.4.1) originates from GR, eqs. (3.4.2) from QFT.
Equations (3.4.3) realize the indeterministic conception.
Equations (3.4.5) define cosmic time $t$ and the family
$\{S_t\}$.

Note that eq.~(3.4.3$a$) may be obtained in the same style as
eq.~(3.4.3$b$). Such an approach would correspond to the Zeno
effect.

\subsection{Some features of the theory}

The picture employed in sec. 3.4 differs from the Heisenberg
one in that $\Psi$ changes in accordance with the cosmic energy
determinacy principle.

The "decoherence" in indeterministic quantum gravity has
nothing to do with the decoherence problem in quantum mechanics
(or QFT). In indeterministic quantum gravity, quantum
mechanical decoherence does not exists at all: it would be
connected with the Heisenberg dynamics of observables.

At a quantum jump, $g$ and its first derivatives remain
continuous, and second derivatives and $G$ undergo a jump.
Therefore gravity per se is quantum: $G$ experiences quantum
jumps along with $\Psi$. This proves the name of the theory:
indeterministic quantum gravity.

The operator $T$ of the energy-momentum tensor may be connected
with the family $F$ of quantum fields, and eqs. (3.4.2) may be
connected with the equations for the fields.

Any $\Psi$ may serve as an eigenvector of $H_t$: this
requirement is fulfilled by choosing a Hamiltonian. This gives
a condition for $g_t$ as the function of $s\in S_t$. In the
local inertial frame approximation with the Hamiltonian
$H_{inert}$, $\Psi$, in general, is not an eigenvector of
$H_{inert}$.

\subsection{QFT and GR approximations}

QFT and GR are obvious approximations of indeterministic
quantum gravity.

Neglecting gravity, we obtain from eq.~(3.4.1)
$$g=\eta  \eqno{(3.6.1)}$$
--- the Minkowski metric. Now there is no privileged reference
frame. In any inertial reference frame, $H$ is time
independent, so that eq.~(3.4.3$a$) leads to
$$\Psi = const,  \eqno{(3.6.2)}$$
eq.~(3.4.2) reads
$$\frac{dT}{dt}= i[H,T].  \eqno{(3.6.3)}$$
Thus we obtain the Heisenberg picture for matter in the
Minkowski spacetime. There are no quantum jumps: the dynamics
of QFT is deterministic. Thus, without gravity there is no
indeterminism.

Neglecting quantum properties of matter, we obtain from
eq.~(3.4.1) the classical Einstein equation, which corresponds
to GR. It is deterministic theory.

Thus indeterminism results only from the combination of quantum
and gravitational properties of the universe.

\section{The effect of instantaneous transference}

Quantum jumps described by eq.~(3.4.3$b$) imply the effect of
instantaneous transference. As an example, a situation of type
occurring in the Einstein-Podolsky-Rosen problem may serve. Let
the wave function of an electron be split into two parts
widely separated, and let the reduction of the wave function
occur due to an interaction involving one of the parts only.
Then the reduction of the second part occurs simultaneously
with that of the first part. Needless to say the simultaneity
refers to the cosmic time $t$.

Since the change of $\Psi_t$ (3.4.3$b$) results in changing
the metric $g$, a possibility for transferring signals
instantaneously arises. A transmitter acts on the first part
of the wave function, and a receiver is a system near the
second part which does not interact with the electron and is
sensitive to the change of $g$.

It cannot be too highly stressed that the effect of
instantaneous transference by no means contradicts QFT, GR, and
the causality principle. As for QFT and GR, the effect is
beyond their scope, being especially indeterministic one. As
to the causality principle,  the simultaneity refers to the
cosmic reference frame only. What is more, the effect assigns
a direct physical meaning to the concept of the cosmic time.

Note that to have a possibility to transmit an instantaneous
signal, one has to previously transfer a material excitation,
whose speed is not greater than that of light.

\section{Some possible ways to develop the theory}

The conceptual structure of indeterministic quantum gravity
expounded in sec. 3.4 is fairly simple. But as to concrete
problems, the mathematical equations of the theory are
extremely involved. Therefore, it is important to direct ways
to develop the theory. The most important problem concerns
physical reasons for level branching.

\subsection{Particle transformations and level branching}

Consider a transformation of particles in the Robertson-Walker
spacetime. The energy of a free particle in the cosmic
reference frame is
$$\varepsilon=\left(E^2+\frac{b^2}{a^2}\right)^{1/2}, a=a(t), b=const,
\eqno{(5.1.1)}$$
where $E$ is the rest energy. Let the state vector of a system
of particles be
$$\psi=c_1\psi_1+c_2\psi_2,  \eqno{(5.1.2)}$$
where $\psi_1$ and $\psi_2$ relate to different collections of
particles. Let energy levels corresponding to $\psi_1$ and
$\psi_2$ coincide for some $t$. With change of $t$, the level
for $\psi$ branches.

Thus the particle transformation results in the level branching
and, by the same token, in a quantum jump.

\subsection{Exponential decay}

Deterministic time evolution for a quantum process is
determined by eqs.~(3.4.2$a$),(3.4.3$a$) or, in the QFT
approximation, eqs.~(3.6.3),(3.6.2). Usually, the process is
described in terms of the Schr\"odinger picture. Let $c(t)$
be the amplitude of the initial state;
$$|c(t)|^2=e^{-t/\tau}  \eqno{(5.2.1)}$$
holds. The problem is that of quantum jump from the initial
state into final one. Let $t_i, i=1,2,\ldots,$ be moments of
branching points. Then the probability to retain the initial
state is
$$w(t_n+0)=e^{-t_1/\tau}\cdot e^{-(t_2-t_1)/\tau}\cdots
e^{-(t_n-t_{n-1})/\tau}=e^{-t_n/\tau},  \eqno{(5.2.2)}$$
which describes the exponential decay. We underline the
discreteness of the set of branching points, which eliminates
the Zeno effect, i.e., retaining the initial state. (This
effect does take place for the continuous evolution described
by eq.~(3.4.3$a$).)

\section{Quantum measurement}

One of the most important manifestations of indeterminism is
quantum measurement. The problem of the latter is one
of the most difficult ones [12--14,19,11]. Indeterministic
quantum gravity opens a new avenue of attack on this problem.

In particular, the results of sec. 5.1 may be employed. For
example, the measurement of the position of an electron by the
use of a luminescent screen involves creating a photon.

\section{The eternal universe}

Cosmology is inherent in the formulation of indeterministic
quantum gravity: $\Psi_t$ and $H_t$ relate to the universe as
a whole. On the other hand, this theory may be used for solving
the cosmological problem of the genesis of the universe. This
problem is avoided in the oscillating model of the universe.
But it is generally taken that this model faces one severe
theoretical difficulty [20]. In each cycle the entropy is
increased by a kind of friction as the universe expands and
contracts. The increase of the entropy results in that of the
minimal value of the energy. For the present cycle this value
is not infinite, so it is hard to see how the universe could
have experienced an infinite number of cycles.

Indeterministic quantum gravity allows to overcome the
difficulty described. In this theory the minimal value of the
energy is an eigenvalue of the Hamiltonian $H_t$, and any
eigenvalue is finite (excluding $t$ of cosmic singularity).

\end{document}